\documentstyle[12pt]{article}
\makeatletter
\@addtoreset{equation}{section}
\makeatother

\def\lb#1{\label{#1}}
\def\l#1{\lb{#1}}
\def\r#1{(\ref{#1})}
\def\c#1{\cite{#1}}
\def\i#1{\bibitem{#1}}
\def\beq{\begin{equation}}
\def\eeq{\end{equation}}
\def\bez{\begin{displaymath}}
\def\eez{\end{displaymath}}
\def\beb#1\l#2\eeb{\begin{equation}
\begin{array}{c} #1 \qquad
\end{array} \label#2  \end{equation}}
\def\bey#1\eey{\begin{displaymath}
\begin{array}{c} #1  \end{array}  \end{displaymath}}

\begin{document}

\begin{titlepage}
\begin{flushright}
December, 31, 2001\\
math-ph/0112062
\end{flushright}

\begin{centering}
\vfill
{\bf 
Group Transformations of Semiclassical Gauge Systems
}
\vspace{1cm}

O. Yu. Shvedov \footnote{shvedov@qs.phys.msu.su} \\
\vspace{0.3cm}
{\small {\em Sub-Dept. of Quantum Statistics and Field Theory,}}\\
{\small{\em Department of Physics, Moscow State University }}\\
{\small{\em Vorobievy gory, Moscow 119899, Russia}}

\vspace{0.7cm}

\end{centering}

{\bf Abstract}

Semiclassical systems  being symmetric under Lie group are studied.  A
state of a semiclassical system may be viewed as a set  $(X,f)$  of  a
classical state  $X$ and a quantum state $f$ in the external classical
background $X$.  Therefore, the set of all semiclassical states may be
considered as  a bundle ("semiclassical bundle").  Its base $\{X\}$ is
the set of all classical  states,  while a  fibre is a  Hilbert  space
${\cal F}_X$ of quantum states in the external background $X$. Symmetry
transformation of  a  semiclassical  system  may  be  viewed   as   an
automorphism of  the semiclassical bundle.  Automorphism groups can be
investigated with  the  help  of  sections  of  the  bundle:  to   any
automorphism of  the bundle one assigns a transformation of section of
the bundle.  Infinitesimal properties of transformations  of  sections
are investigated;  correspondence  between Lie groups and Lie algebras
is discussed.  For gauge theories,  some points of  the  semiclassical
bundle are  identified:  a  gauge  group acts on the bundle.  For this
case, only gauge-invariant sections of the  semiclassical  bundle  are
taken into account.

\vspace{0.7cm}

\vfill \vfill
\noindent

\end{titlepage}
\newpage
\sloppy

\section{Introduction}

Semiclassical approximation  is  a   very   important   technique   of
constructing asymptotic  solutions  of  differential  equations with a
small parameter ${\varepsilon}$:
\beq
i{\varepsilon} \frac{\partial\psi}{\partial    t}     =     H(x,     -
i{\varepsilon}\frac{\partial}{\partial x})  \psi,  \qquad  x  \in {\bf
R}^n, t\in {\bf R}.
\l{1}
\eeq
There are a lot of semiclassical ansatzes that  approximately  satisfy
eq.\r{1} as  ${\varepsilon}\to 0$:  the well-known WKB substitution is
not the only possible semiclassical wave function. The Maslov complex
WKB theory \c{M1} tells us that the wave function of the type
\beq
\psi^t(x) =  e^{\frac{i}{{\varepsilon}}S^t} e^{\frac{i}{{\varepsilon}}
P^t(x-Q^t)} f^t(\frac{x-Q^t}{\sqrt{{\varepsilon}}})
\l{2}
\eeq
also satisfies approximately eq.\r{1}. Here $S^t\in {\bf R}$, $P^t,Q^t
\in {\bf R}^n$,  $f^t(\xi)$ is a smooth rapidly  damping  function  as
$\xi\to \infty$.  The quantum state \r{2} corresponds to the classical
particle with  coordinates  $Q^t$  and  momenta  $P^t$.  The  function
$f^t(\xi)$ specifies   the  shape  of  the  wave  packet.  It  can  be
interpreted as a quantum state in the external  classical  background
$(S^t,P^t,Q^t)$.

Other semiclassical  wave  functions  (including  WKB  states)  may be
viewed as superpositions of states \r{2} \c{MS}.

At the fixed moment of time,  a semiclassical state of the type  \r{2}
may be  viewed as a set $(X,f)$,  where $X=(S,P,Q) \in {\bf R}^{2n+1}$
may be considered as a classical state, while the wave function $f$ is
a quantum  state in a given external classical background $X$.  Set of
all semiclassical states of the form \r{2} may  be  interpreted  as  a
bundle ("semiclassical bundle" \c{S1,S2}).  Its base is ${\cal X} = \{
X \} = {\bf R}^{2n+1}$,  while fibres ${\cal F}_X$ are Hilbert  spaces
$L^2({\bf R}^n)$. Point $(X \in {\cal X}, f \in {\cal F}_X)$ is then a
semiclassical state.

Substitution of expression \r{2} to eq.\r{1} gives  us  the  following
relations \c{M1,MS}
\beq
\dot{S}^t = P^t \dot{Q}^t - H(Q^t,P^t), \qquad
\dot{Q}^t = \frac{\partial H}{\partial P}(Q^t,P^t);
\qquad
\dot{P}^t = - \frac{\partial H}{\partial Q}(Q^t,P^t).
\l{3a}
\eeq
\beq
i \dot{f}^t(\xi) = \left[
\frac{1}{2} \xi \frac{\partial^2 H}{\partial Q \partial Q} \xi
+ \xi     \frac{\partial^2     H}{\partial     Q      \partial      P}
(-i\frac{\partial}{\partial\xi})
+ (-i\frac{\partial}{\partial\xi})
\frac{\partial^2H}{\partial P\partial P}
(-i\frac{\partial}{\partial\xi})
\right] f^t(\xi).
\l{3b}
\eeq
Therefore, the evolution transformations
\beb
u_t: (S^0,P^0,Q^0) \in {\cal X} \mapsto (S^t,P^t,Q^t) \in {\cal X},\\
U_t(u_tX \gets X): f^0 \in {\cal F}_X \mapsto f^t \in {\cal F}_{u_tX}
\l{4}
\eeb
are specified.  The set of transformations \r{4} is an automorphism  of
the semiclassical bundle.

A conception  of  semiclassical  bundle  may  be  introduced  also  in
statistical mechanics (${\varepsilon}$ is $1/N$, where $N$ is a number
of particles),  quantum  field  theory  (${\varepsilon}$ is a coupling
constant) \c{MS}, for quantum gauge theories \c{Sg}.
For these  cases,  the semiclassical bundle is not trivial but locally
trivial.

In this paper the properties of semiclassical bundles being  invariant
under Lie groups are investigated.  Let $({\cal Z}, {\cal X}, \pi)$ be
a semiclassical bundle ($\cal Z$ be a space of the bundle, $\pi: {\cal
Z} \to {\cal X}$ be the projection). Let us start from definitions.

{\bf Definition 1.1.} {\it We say that a Lie group $G$ {\sf acts} on a
semiclassical bundle $({\cal Z}, {\cal X},\pi)$ if:\\
(a) for any $g\in G$ the transformation $u_g:  {\cal X} \to {\cal  X}$
such that \\
(i) the mapping $(g,X) \mapsto u_gX$ is smooth;\\
(ii) $u_{g_1g_2} = u_{g_1} u_{g_2}$;\\
is specified;\\
(b) for  any  $g\in  G$  and  $X  \in  {\cal X}$ the unitary operators
$U_g(u_gX \gets X): {\cal F}_X \to {\cal F}_{u_gX}$ such that\\
(i) the  mapping  $(g,h)  \mapsto U_g(u_hX \gets u_{g^{-1}} u_h X)$ is
strongly continuous for any $X \in {\cal X}$;\\
(ii)
\beq
U_{g_1} (u_{g_1}  u_{g_2} X \gets u_{g_2}X) U_{g_2} (u_{g_2}X \gets X)
= U_{g_1g_2} (u_{g_1g_2}X \gets X)
\l{5}
\eeq
are given.
}

Composition law \r{5} is not  suitable  for  investigations.  However,
formula \r{5}  can be taken to the group form.  Consider the following
transformations of sections of the bundle $({\cal Z}, {\cal X}, \pi)$.
To each  element  $g\in  G$ assign the mapping ${\cal U}_g$ that takes
each section $\Psi$ of the bundle $({\cal Z},{\cal X},\pi)$ of the form
\bez
\{ \Psi_X \in {\cal F}_X, X\in {\cal X}\}
\eez
to the section ${\cal U}_g\Psi$ of the form
\bez
({\cal U}_g \Psi)_{u_gX} = U_g(u_gX \gets X) \Psi_X.
\eez
The transformation ${\cal U}_g$ obey the group law
\bez
{\cal U}_{g_1} {\cal U}_{g_2} = {\cal U}_{g_1g_2}.
\eez
Under certain  conditions,  the  operators  $U_g(u_gX\gets   X)$   are
uniquely determined  by  the  transformations  ${\cal  U}_g$  and wise
versa. Instead of sections of the bundle $({\cal Z},  {\cal X}, \pi)$,
one can consider gauge orbits on the base
\bez
[\overline{X}] = \{ u_h\overline{X}, h\in G\}
\eez
and sections    of    the    subbundles   $(\pi^{-1}   [\overline{X}];
[\overline{X}], \pi)$     with     $\pi     \equiv      \pi|_{\pi^{-1}
[\overline{X}]}$. These problems are to be considered in section 2.

Analogous ideas   were   used   in   \c{U1}.  Properties  of  "unitary
propagators" being unitary transformations  $U(t,\tau)$,  $t,\tau  \in
{\bf R}$ of the Hilbert space $\cal H$ such that
\bez
U(t_1,t_2) U(t_2,t_3) = U(t_1,t_3)
\eez
were investigated. The one-parametric group of unitary transformations
${\cal V}(t)$ of the Hilbert space $L_2({\bf R} \to {\cal H})$ of  the
form
\bez
({\cal V}(t) f)(\tau) = U(t,t-\tau) f(t-\tau),  \qquad f:  {\bf R} \to
{\cal H}.
\eez
were assigned to each unitary propagator $U(t,\tau)$.  It appeared  to
be more  convenient  to  investigate properties of the transformations
${\cal V}(t)$ rather than  of  the  propagator  $U(t,\tau)$.  For  our
purposes, it  is  more  suitable  to  consider  the  Banach  space  of
continuous vector functions $f$ with respect to the norm $\sup ||f||$.

For constructing group representations and checking the group law,  it
is convenient   to   consider   infinitesimal   transformations.   One
investigates then Lie algebras instead of Lie groups. In section 3, the
representation of the Lie algebra of the group $G$ in the Garding-type
spaces \c{BR,Gard} is assigned to the representation ${\cal  U}_g$  of
the Lie group (i.e. to the set of operators $U_g(u_gX \gets X)$).

The inverse  problem of integrating representations of Lie algebras is
nontrivial \c{BR,Nelson,Flato}.   In   section   4,   the   sufficient
conditions of  integrability  of Lie algebra representations in spaces
of sections  are  presented.  For  example,   they   are   useful   in
investigating the  property  of  Poincare  invariance of semiclassical
field theory \c{Shvedov}.

Gauge-invariant systems are widely considered in quantum field theory.
In the semiclassical approximation,  the principle of gauge invariance
means that some equivalence relation should be introduced in the space
of the semiclassical bundle \c{Sg}.

{\bf Definition 1.2.} {1.  \it We say that a {\sf gauge} group $\cal L$ {\sf
acts} on a semiclassical bundle $({\cal Z}, {\cal X}, \pi)$ if for any
$\alpha\in {\cal L}$ the set
\bez
\lambda_{\alpha}: {\cal X} \to {\cal X}; \qquad
V_{\alpha}(\lambda_{\alpha}X \gets   X):   {\cal   F}_X   \to    {\cal
F}_{\lambda_{\alpha} X}
\eez
of a  smooth  mapping  $\lambda_{\alpha}$  and unitary transformations
$V_{\alpha}(\lambda_{\alpha} X \gets X)$ such that\\
(i) the  set $(\lambda_{\alpha},V_{\alpha})$ specifies an automorphism
of the bundle $({\cal Z}, {\cal X}, \pi)$;\\
(ii)
\bez
\lambda_{\alpha_1} \lambda_{\alpha_2}   =  \lambda_{\alpha_1\alpha_2};
\qquad
V_{\alpha_1}(\lambda_{\alpha_1} \lambda_{\alpha_2}       X       \gets
\lambda_{\alpha_2} X)
V_{\alpha_2}(\lambda_{\alpha_2} X \gets X)
= V_{\alpha_1\alpha_2}(\lambda_{\alpha_1} \lambda_{\alpha_2} X \gets X)
\eez
is given. \\
2. Two  points  $(X_1,f_1),  (X_2,f_2)  \in  {\cal Z}$ are called {\sf
gauge-equivalent}, $(X_1,f_1) \sim (X_2,f_2)$,  if for some $\alpha\in
{\cal L}$
\bez
\lambda_{\alpha} X_1 = X_2; \qquad V_{\alpha} (X_2\gets X_1) f_1 = f_2.
\eez
}

For gauge  systems,  it  is  not  required  for  property  \r{5} to be
satisfied. A weaker condition is postulated: points
\bez
(u_{g_1}u_{g_2} X, U_{g_1}(u_{g_1}u_{g_2} X \gets u_{g_2}X)
U_{g_2}(u_{g_2}X \gets X) f)
\qquad {\sf and} \qquad
(u_{g_1g_2} X, U_{g_1g_2}(u_{g_1}u_{g_2} X \gets X) f)
\eez
are required to be gauge-equivalent. We also require that
\bez
(u_gX_1, U_g(u_gX_1 \gets X_1) f_1) \sim
(u_gX_2, U_g(u_gX_2\gets X_2) f_2),
\eez
provided that $(X_1,f_1) \sim (X_2,f_2)$.

For gauge-invariant  systems,  actions   of   groups   can   be   also
investigated by  the section method.  To satisfy the group law for the
operators ${\cal U}_g$, spaces of gauge-invariant sections obeying the
condition
\bez
\Psi_{\lambda_{\alpha}X} =   V_{\alpha}  (\lambda_{\alpha}X  \gets  X)
\Psi_X
\eez
should be considered.  Analogs of results of sections 2-4 are obtained
for the gauge-invariant systems in section 5.

\section{Transformations of sections of a semiclassical bundle}

{\bf 1.} Let $\overline{X} \in {\cal X}$.
By $({\cal   Z}_{[\overline{X}]},[\overline{X}],\pi)$  we  denote  the
subbundle of the bundle $({\cal Z},{\cal X},\pi)$ such that
\bez
[\overline{X}] = \{u_g\overline{X}, g\in G\}.
\eez
Let $G_0(\overline{X})$ be the stationary subgroup of the point
$\overline{X}$ (set of all $g_0\in G$ such that $u_{g_0}\overline{X} =
\overline{X}$). By $G/G_0(\overline{X})$ we denote the factorspace  of
classes $gG_0$.  To  each strongly continuous section $\{X \mapsto
\Psi_X,  X\in [X]\}$ of the subbundle
$({\cal Z}_{[\overline{X}]},[\overline{X}],\pi)$  assign  the strongly
continuous vector function $\Psi(g) = \Psi_{u_gX}$,  which is constant
on each class $gG_0$.  Therefore,  $\Psi(g)$ can be viewed as a vector
function on $G/G_0(\overline{X})$.  If the support of this function is
compact, then  we  say  that  the section $\Psi$ is {\sf of the class}
$C_0([\overline{X}])$. By definition, put
\beq
||\Psi||_{[\overline{X}]} = \max_g ||\Psi_{u_g\overline{X}}||.
\l{6}
\eeq
By $C([\overline{X}])$  we denote the closure of $C_0([\overline{X}])$
with respect to the norm \r{6}. Consider the following operator
\bez
({\cal U}_g^{[\overline{X}]} \Psi)_{u_h\overline{X}} = U_g(u_h
\overline{X} \gets u_{g^{-1}h}\overline{X}) \Psi_{u_{g^{-1}h}
\overline{X}}
\l{7}
\eez
in the space of sections of the subbundle $({\cal Z}_{[\overline{X}]},
[\overline{X}],\pi)$. It happens  that  properties  of  the  operators
$U_g(X\gets u_{g^{-1}}X)$  can  be  reformulated  in  terms  of  ${\cal
U}_g^{[\overline{X}]}$.

{\bf Lemma 2.1.} {\it Let $\Psi \in C_0([\overline{X}])$.  Then ${\cal
U}_g^{[\overline{X}]}\Psi \in C_0([\overline{X}])$.  }

{\bf Proof.}   The   mapping   $(g,X)  \mapsto  u_gX$  is  continuous.
Therefore, the support of ${\cal  U}_g^{[\overline{X}]}\Psi$  is  compact.
Let us  check  that  ${\cal U}_g^{[\overline{X}]} \Psi$ is continuous.
The property of local triviality of vector  bundle  implies  that  all
spaces ${\cal F}_X$ may be viewed as identical.  Let $h_n \to h$ be an
arbitrary sequence. One has:
\bey
||({\cal U}_g^{[\overline{X}]} \Psi)_{u_{h_n}X} -
({\cal U}_g^{[\overline{X}]} \Psi)_{u_{h}X}|| =\\
||U_g(u_{h_n}X \gets u_{g^{-1}h_n}X) \Psi_{u_{g^{-1}h_n}X}
- U_g(u_{h}X \gets u_{g^{-1}h}X) \Psi_{u_{g^{-1}h}X}|| \le
\\
||U_g(u_{h_n}X \gets u_{g^{-1}h_n}X)|| ||\Psi_{u_{g^{-1}h_n}X} -
\Psi_{u_{g^{-1}h}X}|| \\
+  ||(U_g(u_{h_n}X   \gets   u_{g^{-1}h_n}X)   -
U_g(u_hX \gets  u_{g^{-1}h}X)) \Psi_{u_{g^{-1}h}X} || \to_{n\to\infty}
0.
\eey
Lemma 2.1 is proved.

{\bf Lemma 2.2.} {\it Let $\Psi \in C_0([\overline{X}])$. Then
$||{\cal U}_g^{[\overline{X}]}\Psi||_{[\overline{X}]}                =
||\Psi||_{[\overline{X}]}$.
}

The proof is straightforward.

Therefore, the operators ${\cal U}_g^{[\overline{X}]}$ can be uniquely
extended to  the  space  $C(\overline{X})$.  The  isometric   property
remains valid.

Property \r{5} can be reformulated as follows.

{\bf Lemma 2.3.} {\it
For the operators ${\cal U}_g^{[\overline{X}]}$,  the group properties
are satisfied:
\beq
{\cal U}_{g_1}^{[\overline{X}]}
{\cal U}_{g_2}^{[\overline{X}]} =
{\cal U}_{g_1g_2}^{[\overline{X}]}, \qquad
{\cal U}_e^{[\overline{X}]} = 1.
\l{7a}
\eeq
}

The proof is trivial.

{\bf Lemma 2.4.} {\it
The operator ${\cal U}_g^{[\overline{X}]}$ is strongly continuous with
respect to $g$ at the point $g=e$.
}

{\bf Proof.}  Let  $\Psi  \in   C_0([\overline{X}])$.   Consider   the
expression
\bey
||{\cal U}_g^{[\overline{X}]}  \Psi - \Psi||_{[\overline{X}]} = \sup_h
||({\cal U}_g^{[\overline{X}]}        \Psi)_{u_h\overline{X}}        -
\Psi_{u_h\overline{X}}|| =    \sup_h    ||U_g(u_h\overline{X}    \gets
u_{g^{-1}h} \overline{X})    \Psi_{u_{g^{-1}h}     \overline{X}}     -
\Psi_{u_h\overline{X}}||.
\eey
Let us  prove that it tends to zero as $g\to e$.  Assume the converse.
Then for some sequences of elements $g_n \in G$ such that  $g_n\to  e$
and of equivalence classes $\overline{h}_n \in G/G_0$
\beq
||U_{g_n}(u_{\overline{h}_n} \overline{X}       \gets      u_{g_n^{-1}}
u_{\overline{h}_n} \overline{X})                    \Psi_{u_{g_n^{-1}}
u_{\overline{h}_n}X} - \Psi_{u_{\overline{h_n}}X}||\ge C_1>0.
\l{8}
\eeq
Let $U\subset G$ be an open set such that $e\in U$ and  $\overline{U}$
is compact.  By  $D$  we  denote  the  set  of  all  $\overline{h} \in
G/G_0(\overline{X})$ such that $\Psi_{u_{\overline{h}}X} \ne 0$.  Then
$\overline{D}$ is     compact.     There    exists    a    subsequence
$\{\overline{h_{n_k}}\}$    such   that    $\overline{h_{n_k}}    \in
\overline{U} \overline{D}$.  Otherwise,  the left-hand side of formula
\r{8} will be zero for sufficiently large $n$. Since the sequence
$\overline{h_{n_k}}$ belongs   to   the   compact   set  $\overline{U}
\overline{D}$., one     can     choose     a      subsequence
$\{ \overline{h_{n_{k_l}}} \}$
being  convergent  with  respect  to  the
$G/G_0(\overline{X})$-topology. Therefore, for some representatives
$h_{l} \in \overline{h_{n_{k_l}}}$ one has $h_l \to h$ with respect to
the $G$-topology.  Denote $g_l' = g_{n_{k_l}}$.  Formula \r{8} implies
that
\bez
||U_{g_l'} (u_{h_l}     \overline{X}     \gets      u_{g_l'{}^{-1}h_l}
\overline{X}) \Psi_{u_{g_l'{}^{-1}h_l}\overline{X}} -\Psi_{u_{h_l}
\overline{X}}||\ge C_1.
\eez
If one combines tis with eq.\r{5}, one gets
\beb
||
U_{h_l^{-1}g_l'} (\overline{X} \gets u_{g_l'{}^{-1}h_l} \overline{X})
\Psi_{u_{g_l'{}^{-1}h_l} \overline{X}} -
U_{h_l^{-1}} (\overline{X} \gets u_{h_l} \overline{X})
\Psi_{u_{h_l} \overline{X}}|| \ge C_1.
\l{9}
\eeb
Since the mapping
\bez
h \mapsto U_{h^{-1}}(\overline{X}  \gets u_h\overline{X})
\eez
is strongly continuous, while the operators
$U_{h^{-1}}(\overline{X} \gets  u_h\overline{X})$  are  unitary,   one
obtains:
\bey
||U_{h_l^{-1}}(\overline{X}  \gets u_{h_l} \overline{X})
\Psi_{u_{h_l} \overline{X}} -
U_{h^{-1}}(\overline{X}  \gets u_{h} \overline{X})
\Psi_{u_{h} \overline{X}}|| \to 0;\\
||U_{h_l^{-1}g_l'} (\overline{X} \gets u_{g_l'{}^{-1}h_l}\overline{X})
\Psi_{u_{g_l'{}^{-1}h_l}\overline{X}} -
U_{h^{-1}}(\overline{X}  \gets u_{h} \overline{X})
\Psi_{u_{h} \overline{X}}|| \to 0
\eey
as $l \to \infty$. This contradicts \r{9}. Hence, the unitary operator
${\cal U}_g^{[\overline{X}]}$  tends to 1 in the strong sense on the set
$C_0([\overline{X}])$,
\beq
\lim_{g\to e}      ||{\cal      U}_g^{[\overline{X}]}      \Psi      -
\Psi||_{[\overline{X}]} = 0, \qquad
\Psi \in C_0([\overline{X}]).
\l{9a}
\eeq
However, the     set     $C_0([\overline{X}])$     is     dense     in
$C([\overline{X}])$. Then  the  Banach-Steinhaus  theorem  (see,   for
example, \c{KA})  implies  that property \r{9a} is valid for $\Psi \in
C([\overline{X}])$ as well. Lemma is proved.

Let $\alpha_X$,  $X\in {\cal  X}$  be  a  complex  continuous  bounded
function on     ${\cal     X}$.     By    $v_{[\overline{X}]}[\alpha]:
C_0([\overline{X}]) \to C_0([\overline{X}])$ denote the multiplication
operator of the form
\bez
(v_{[\overline{X}]}[\alpha] \Psi)_{u_h\overline{X}}                  =
\alpha_{u_h\overline{X}} \Psi_{u_h\overline{X}}.
\eez
The operator   is   bounded.   It   can   be    extended    then    to
$C([\overline{X}])$ uniquely. Denote also
\bez
(w_g\alpha)_X = \alpha_{u_{g^{-1}}X}.
\eez

{\bf Lemma 2.5.} {\it For $\Psi \in C([\overline{X}])$, one has
\beq
{\cal U}_g^{[\overline{X}]}    v_{[\overline{X}]}[\alpha]    \Psi    =
v_{[\overline{X}]} [w_g\alpha] {\cal U}_g^{[\overline{X}]} \Psi.
\l{10}
\eeq
}

{\bf Proof.} For $\Psi \in C_0([\overline{X}])$, check of eq.\r{10} is
trivial. Relation \r{10} can be extended to $C([\overline{X}])$, since
operators                             $v_{[\overline{X}]}[w_g\alpha]$,
$v_{[\overline{X}]}[\alpha]$, ${\cal     U}_g^{[\overline{X}]}$    are
continuous. Lemma is proved.

Combining Lemmas 2.1 - 2.5, one gets

{\bf Theorem 2.1.} {\it
Suppose that  the  group  $G$ acts on the semiclassical bundle $({\cal
Z},{\cal X},\pi)$.  Then the operators  ${\cal  U}_g^{[\overline{X}]}:
C([\overline{X}]) \to C([\overline{X}])$ of the form \r{7}:\\
- are   invariant   under   transformation   $\overline{X}   \to   u_h
\overline{X}$;\\
- are isometric;\\
- satisfy properties \r{7a} and \r{10};\\
- are strongly continuous with respect to $g$ at $g=e$.
}

{\bf 2.} The inverse theorem is also valid.

{\bf Theorem   2.2.}   {\it   Suppose   that  for  any  $g\in  G$  the
transformation $u_g: {\cal X} \to {\cal X}$ such that:\\
(i) the mapping $(g,X) \mapsto u_gX$ is smooth;\\
(ii) $u_{g_1g_2} = u_{g_1}u_{g_2}$;\\
is specified.  Let  also  for any $g\in G$ and $\overline{X} \in {\cal
X}$ the operators ${\cal U}_g^{[\overline{X}]}:  C([\overline{X}]) \to
C([\overline{X}])$ such that:\\
(i) ${\cal U}_g^{[\overline{X}]}$ is invariant under transformation
$\overline{X} \mapsto u_h \overline{X}$;\\
(ii) ${\cal U}_g^{[\overline{X}]}$ is isometric;\\
(iii) properties \r{7a} and \r{10} are satisfied;\\
(iv) ${\cal U}_g^{[\overline{X}]}$ is strongly continuous with respect
to $g$ at $g=e$\\
is given. The for all $g\in G$ and $X \in {\cal X}$ there exist unique
unitary operators  $U_g(u_gX \gets X):  {\cal F}_X \to {cal F}_{u_gX}$
such that:\\
(i) the  mapping  $(g,h)  \mapsto U_g(u_hX \gets u_{g^{-1}} u_h X)$ is
strongly continuous for any $X\in {\cal X}$;\\
(ii) the property \r{5} is satisfied;\\
(iii) the operator ${\cal U}_g^{[\overline{X}]}$ has the form \r{7}.
}

To prove Theorem 2.2, we need several Lemmas.

{\bf Lemma 2.6.} {\it
The operator $U_g(X\gets u_{g^{-1}}X)$ is uniquely found from relation
\r{7}.
}

{\bf Proof.} Namely, let $X=u_h\overline{X}$, $\Psi_0$ be some element
of ${\cal    F}_{u_{g^{-1}}X}$.    Choose    a   section   $\Psi   \in
C_0([\overline{X}])$ of  the  subbundle  $({\cal  Z}_{[\overline{X}]},
\overline{X}, \pi)$ such that $\Psi_{u_{g^{-1}}X}= \Psi_0$. Set
\beq
U_g(X\gets u_{g^{-1}}X)  \Psi_0  \equiv  ({\cal  U}_g^{[\overline{X}]}
\Psi)_X.
\l{10a}
\eeq
Show this   procedure   of   constructing   the  operator  $U_g(X\gets
u_{g^{-1}}X)$ to be well defined.
It is  sufficient  to  check  that  $\Psi_{u_{g^{-1}}X}  =  0$ implies
$({\cal U}_g^{[\overline{X}]}    \Psi)_X     =0$.     Suppose     that
$\Psi_{u_{g^{-1}}X} = 0$. This implies that $\Psi_Y$ can be written as
$\Psi_Y = \alpha_Y \Phi_Y$.  Here $\Phi_Y$ is a continuous section  of
$({\cal Z}_{[\overline{X}]},{[\overline{X}]},\pi)$,  $\alpha_Y$  is  a
continuous function     on     ${[\overline{X}]}$       such      that
$\alpha_{u_{g^{-1}}X} = 0$. It is sufficient to set
\bez
\alpha_Y = ||\Psi_Y||^{1/2}, \qquad \Phi_Y = \Psi_Y/||\Psi_Y||^{1/2}.
\eez
One has
\bez
({\cal U}_g^{[\overline{X}]}  \Psi)_X  =  ({\cal U}^{[\overline{X}]}_g
v_{[\overline{X}]}[\alpha] \Phi)_X =  (v_{[\overline{X}]}  [w_g\alpha]
{\cal U}_g^{[\overline{X}]}\Phi)_X   =   \alpha_{u_{g^{-1}}X}   ({\cal
U}_g^{[\overline{X}]}\Phi)_X = 0.
\eez
Therefore, definition \r{10a} is well formulated. Lemma 2.6 is proved.

{\bf Lemma 2.7.} {\it The property \r{5} is satisfied;
$U_e(X\gets X) = 1$.
}

This is a direct corollary of relation \r{7a}.

{\bf Lemma 2.8.} {\it The operator $U_g(u_gX \gets X)$ is
isometric.
}

{\bf Proof.} First, show that
\beq
||U_g(Y\gets u_{g^{-1}}Y) \Phi|| \le ||\Phi||.
\l{10b}
\eeq
Choose a      section     $\Psi$     of     the     bundle     $({\cal
Z}_{[\overline{X}]},{[\overline{X}]},\pi)$ such that
\bez
\max_{X\in {[\overline{X}]}} ||\Psi_X|| = ||\Psi_{u_{g^{-1}}Y}||.
\eez
Then
\bez
||({\cal U}_g^{[\overline{X}]}       \Psi)_Y||       \le       ||{\cal
U}_g^{[\overline{X}]} \Psi||_{[\overline{X}]}                        =
||\Psi||_{[\overline{X}]} = ||\Psi_{u_{g^{-1}}Y}||.
\eez
Hence, property \r{10} is satisfied. Besides, Lemma 2.7 implies
\beq
U_g(Y\gets u_{g^{-1}}Y) U_{g^{-1}} (u_{g^{-1}}Y \gets Y) = 1
\l{11}
\eeq
and
\bez
||\Phi|| =
||U_g(Y\gets u_{g^{-1}}Y) U_{g^{-1}} (u_{g^{-1}}Y \gets Y) \Phi|| \le
||U_{g^{-1}} (u_{g^{-1}}Y \gets Y) \Phi|| \le ||\Phi||
\eez
for any  $\Phi  \in  {\cal  F}_Y$.  Thus,  the inequality signs can be
substituted by equality signs. The operators
$U_g(Y\gets u_{g^{-1}}Y)$ are isometric then. Lemma 2.8 is proved.

Combining Lemmas  2.7 and 2.8,  we find that the operators $U_g(Y\gets
u_{g^{-1}}Y)$ are unitary.

{\bf Lemma 2.9.} {\it Let the operators ${\cal U}_g^{[\overline{X}]} :
C({[\overline{X}]}) \to  C({[\overline{X}]})$  be  strongly continuous
with respect to $g$  at  $g=e$.  Then  the  operator  $U_g(u_hX  \gets
u_{g^{-1}h}X)$ is strongly continuous with respect to $(g,h)$.
}

{\bf Proof.} Let $g_n \to g$, $h_n \to h$ be any convergent sequences.
One has
\bez
U_{g_n} (u_{h_n}X  \gets  u_{g_n^{-1}h_n}X)  = U_{h_nh^{-1}} (u_{h_n}X
\gets u_hX)  U_g(u_hX  \gets  u_{g^{-1}h}X)  U_{g^{-1}h   h_n^{-1}g_n}
(u_{g^{-1}h}X \gets u_{g_n^{-1}h_n}X).
\eez
To prove Lemma 2.9, it is sufficient to check that
\beq
U_{g_n}(u_{g_n}Y \gets Y) \to 1,  \qquad U_{g_n}(Y\gets u_{g_n^{-1}}Y)
\to 1
\l{12}
\eeq
in the  strong  topology.  It is sufficient to prove one of properties
\r{12} because of \r{11}.

Since the  operator  ${\cal   U}^{[\overline{X}]}_{g}$   is   strongly
continuous at $g=e$, one has:
\bez
\sup_{Y=u_hX} ||U_{g_n}(Y\gets  u_{g_n^{-1}}Y)  \Psi_{u_{g_n^{-1}}Y} -
\Psi_Y|| \to 0
\eez
as $g\to e$. Therefore, property
\bez
||U_{g_n} (Y\gets u_{g_n^{-1}}Y) \Psi_{u_{g_n^{-1}}Y} - \Psi_Y|| \to 0
\eez
is valid for all $Y$.  Let us choose a section $\Psi$ such that it  is
constant in  some vicinity of the point $Y$ for some trivialization of
the semiclassical bundle.  One gets then property \r{12}.  Lemma 2.9 is
proved.

Thus, all statements of Theorem 2.2 are checked.

{\bf 3.} Let us introduce the function-valued bilinear form that takes
each set $(\Phi \in C({[\overline{X}]}),\Psi \in C({[\overline{X}]}))$
to the complex function on ${[\overline{X}]}$ of the form
\bez
<\Phi,\Psi>_X = (\Phi_X,\Psi_X), \qquad X \in {[\overline{X}]}.
\eez
The bilinear form resembles inner product. It satisfies the properties:
\bey
<\Phi,\Psi_1+\Psi_2>_X = <\Phi,\Psi_1>_X + <\Phi,\Psi_2>_X;\\
<\Phi,\Psi>_X^* = <\Psi,\Phi>_X;\\
<\Phi,v_{[\overline{X}]}[\alpha]\Psi>_X = \alpha_X <\Phi,\Psi>_X;\\
<\Phi,\Phi>_X \ge 0; \\
<\Phi,\Phi>_X = 0 \qquad for \qquad all \qquad X \equiv \Phi =  0;\\
\sup_X |<\Phi,\Psi>_X| \le ||\Phi|| ||\Psi||;\\
<\Phi,\Psi>_X \qquad is \qquad continuous.
\eey

{\bf Lemma 2.10.} {\it The property
\beq
<{\cal U}_g^{[\overline{X}]} \Phi,  {\cal  U}_g^{[\overline{X}]}>_Y  =
<\Phi,\Psi>_{u_{g^{-1}}Y}
\l{13a}
\eeq
is satisfied.
}

{\bf Proof.} One has
\bey
<{\cal U}^{[\overline{X}]}_g \Phi, {\cal U}_g^{[\overline{X}]} \Psi>_Y
= \\
(U_g(Y \gets u_{g^{-1}}Y) \Phi_{u_{g^{-1}}Y},
U_g(Y \gets u_{g^{-1}}Y) \Psi_{u_{g^{-1}}Y}) =
(\Phi_{u_{g^{-1}}Y},\Psi_{u_{g^{-1}}Y}) =
<\Phi,\Psi>_{u_{g^{-1}}Y}.
\eey

\section{Infinitesimal generators  of semiclassical transformations in
Garding-type spaces}

To study infinitesimal properties of the representation
${\cal U}_g^{[\overline{X}]}$,  it  is  important  to  investigate the
differentiability properties   of    ${\cal    U}_g^{[\overline{X}]}$.
Analogously to \c{Gard},  let us construct a domain $D$ being dense in
$C(\overline{X})$ such   that   ${\cal   U}_g^{[\overline{X}]}$    are
differentiable on $D$.

By $D_0({[\overline{X}]})$  we  denote  the set of the sections of the
subbundle $({\cal Z}_{[\overline{X}]},  {[\overline{X}]},\pi)$ of  the
form
\beq
\Psi_X =    \int    d_Lg    \gamma(g)    U_g(X    \gets   u_{g^{-1}}X)
\Phi_{u_{g^{-1}}X}, \qquad X \in {[\overline{X}]}.
\l{13}
\eeq
Here $d_Lg$  is  the  left-invariant  Haar  measure  on the group $G$,
$\gamma(g)$ is a smooth function on $G$ with  compact  support,  $\Phi
\in C_0({[\overline{X}]})$.

{\bf Lemma 3.1.} {\it
$D_0({[\overline{X}]}) \subset C_0({[\overline{X}]})$.
}

{\bf Proof.}  The  support  of  the  section \r{13} has the form ${\rm
supp} \gamma \quad {\rm supp} \\Phi$ and therefore  compact.  Consider
some local  trivialization  of the bundle $({\cal Z}_{[\overline{X}]},
{[\overline{X}]},\pi)$. Check that section \r{13} is  continuous.  Let
$h_n\to h$. Prove that
\beq
\Psi_{u_{h_n}{[\overline{X}]}} \to \Psi_{u_h{[\overline{X}]}}.
\l{14}
\eeq
One has:
\beb
||\Psi_{u_{h_n}{[\overline{X}]}} -  \Psi_{u_h{[\overline{X}]}}||   \le
\int d_Lg |\gamma(g)| ||\Phi_{u_{g^{-1}h_n}X} - \Phi_{u_{g^{-1}h}X}|| +
\\
\int d_Lg |\gamma(g)|
||(U_g(u_{h_n}X \gets    u_{g^{-1}h_n}X)    -     U_g(u_{h}X     \gets
u_{g^{-1}h}X)) \Phi_{u_{g^{-1}h}X}||.
\l{14a}
\eeb
It follows from the Lesbegue theorem (see,  for example,  \c{KF}) that
the limits  of  integrals  entering  to  eq.\r{14a}  coincide with the
integrals of the limits of integrands.  Hence,  one  obtains  property
\r{14}. Lemma 3.1 is proved.

{\bf Lemma  3.2.}  {\it  Let $h(\tau)$ be a one-parametric subgroup of
$G$, $\Psi \in D_0({[\overline{X}]})$. Then the function
${\cal U}^{[\overline{X}]}_{h(\tau)}       \Psi$       belongs      to
$D_0({[\overline{X}]})$ and is strongly differentiable with respect to
$\tau$ at $\tau=0$. Moreover,
\bez
\frac{d}{d\tau}|_{\tau=0} {\cal U}^{[\overline{X}]}_{h(\tau)} \Psi \in
D_0({[\overline{X}]}).
\eez
}

{\bf Proof.} One has
\bey
({\cal U}_{h(\tau)}^{[\overline{X}]}  \Psi)_X  =  U_{h(\tau)}  (X\gets
u_{h^{-1}(\tau)}X) \Psi_{u_{h^{-1}(\tau)}X} = \\
\int d_Lg  \gamma(g)  U_{h(\tau)g}  (X\gets   u_{g^{-1}h^{-1}(\tau)}X)
\Phi_{u_{g^{-1}h^{-1}(\tau)}X} = \\
\int d_Lg      \gamma(h^{-1}(\tau)g)      U_g(X\gets      u_{g^{-1}}X)
\Phi_{u_{g^{-1}}X}.
\eey
Hence, ${\cal       U}^{[\overline{X}]}_{h(\tau)}       \Psi       \in
D_0({[\overline{X}]})$. Let  us   check   the   property   of   strong
differentiability. One finds:
\beb
(\frac{1}{\tau} ({\cal U}^{[\overline{X}]}_{h(\tau)}-1) \Psi)_X =
\int d_Lg \frac{\gamma(h^{-1}(\tau)g) - \gamma(g)}{\tau}
U_g(X\gets u_{g^{-1}}X) \Phi_{u_{g^{-1}}X} =
\\
\int d_Lg      \int_0^1      d\alpha      \frac{d}{dt}|_{t=\alpha\tau}
\gamma(h^{-1}(t)g) U_g(X\gets u_{g^{-1}}X) \Phi_{u_{g^{-1}}X}.
\l{15a}
\eeb
Formally, this implies
\beq
(\frac{d}{d\tau}|_{\tau=0} {\cal U}_{h(\tau)}^{[\overline{X}]} \Psi)_X
= \int   d_Lg   \frac{d}{dt}|_{t=0}   \gamma(h^{-1}(t)g)    U_g(X\gets
u_{g^{-1}}X) \Phi_{u_{g^{-1}}X}.
\l{15}
\eeq
To justify formula \r{15}, consider the difference of expressions \r{15}
and \r{15a}:
\bey
|| (\frac{1}{\tau} ({\cal U}_{h(\tau)}^{[\overline{X}]} - 1) \Psi -
\frac{d}{d\tau}|_{\tau=0} {\cal          U}_{h(\tau)}^{[\overline{X}]}
\Psi||_{[\overline{X}]} \\
\le \int d_Lg \int_0^1 d\alpha
[\frac{d}{dt}|_{t=\alpha\tau} - \frac{d}{dt}|_{t=0}] \gamma(h^{-1}(t)g))
\sup_{X \in      {[\overline{X}]}}      ||U_g(X\gets      u_{g^{-1}}X)
\Phi_{u_{g^{-1}}X}||.
\eey
It tends to zero as $\tau\to 0$.  Since expression \r{15}  is  of  the
class $D_0({[\overline{X}]})$, all statements of Lemma 3.2 are proved.

{\bf Lemma  3.3.}  {\it  The  set  $D_0({[\overline{X}]})$  is a dense
subset of $C({[\overline{X}]})$.
}

{\bf Proof.}  It  is  sufficient  to  check that any section $\Phi \in
C_0({[\overline{X}]})$ can be approximated by sections \r{13}.  Choose
a sequence $\gamma_n(g)$ such that
\bez
\int d_Lg \gamma_n(g) =1, \qquad \gamma_n(g) \ge 0.
\eez
For any  open  set  $U$  such that $e\in U$ there exists $N$ such that
$supp \gamma_n \subset U$ for all  $n>N$.  Consider  the  sequence  of
sections
\bez
\Psi_X^n =   \int   d_L   g   \gamma_n(g)   U_g(X  \gets  u_{g^{-1}}X)
\Phi_{u_{g^{-1}}X}.
\eez
One has
\bez
||\Psi_X^n -  \Phi_X||  \le   \int   d_Lg   \gamma_n(g)   ||U_g(X\gets
u_{g^{-1}}X) \Phi_{u_{g^{-1}}X} - \Phi_X||.
\eez
Hence,
\bez
||\Psi^n - \Phi||_{[\overline{X}]} \le \int d_L g \gamma_n(g) \sup_h
||U_g(u_{\overline{h}}\overline{X}
\gets
u_{g^{-1}\overline{h}}\overline{X})
\Phi_{u_{g^{-1}\overline{h}}\overline{X}} - \Phi_{\overline{X}}||.
\eez
We have already justified (proof of Lemma 2.4) that
\bez
\sup_{\overline{h}} ||U_g(u_{\overline{h}}\overline{X} \gets
u_{g^{-1}\overline{h}}\overline{X})
\Phi_{u_{g^{-1}\overline{h}}\overline{X}} - \Phi_{\overline{X}}|| \to 0
\eez
as $g\to e$. This means that for any ${\varepsilon}>0$ there exists an
open set $U$ such that $e\in U$ and
\bez
\sup_{\overline{h}} ||U_g(u_{\overline{h}}\overline{X} \gets
u_{g^{-1}\overline{h}}\overline{X})
\Phi_{u_{g^{-1}\overline{h}}\overline{X}} - \Phi_{\overline{X}}|| <
{\varepsilon}
\eez
for all $g \in U$.  However,  for some $N$,  $supp \gamma_n \subset U$
for $n>N$. Therefore,
\bez
||\Psi_X^n - \Phi_X|| < {\varepsilon}.
\eez
Lemma 3.3 is proved.

Let us   extend   the    set    $D_0([\overline{X}])$.    By    ${\cal
C}_0^{\infty}([\overline{X}])$ we denote the set of all smooth complex
functions $\alpha:  X \in [\overline{X}] \mapsto \alpha_X \in {\bf C}$
such that  set of all $\overline{h} \in G/G_0(\overline{X})$ such that
$\alpha_{u_{\overline{h}}X} \ne       0$is       pre-compact        in
$G/G_0(\overline{X})$. By $D([\overline{X}])$ we denote the set of all
linear combinations of the form
\bez
v_{\overline{X}} [\alpha_1] \Psi_1 + ... + v_{\overline{X}} [\alpha_n]
\Psi_n
\eez
with $\alpha_i  \in {\cal C}_0^{\infty}([\overline{X}])$.  $\Psi_i \in
D_0([\overline{X}]$. Notice that $\Psi\in  D([\overline{X}])$  implies
$v_[\overline{X}][\alpha] \Psi  \in D([\overline{X}])$ for $\alpha \in
{\cal C}_0^{\infty}([\overline{X}])$.

{\bf Lemma 3.4.} {\it
Let $h(\tau)$ be a one-parametric subgroup of the Lie group $G$, $\Psi
\in D([\overline{X}])$.      Then       the       section       ${\cal
U}^[\overline{X}]_{h(\tau)} \in    D([\overline{X}])$    is   strongly
differentiable with respect to $\tau$ at $\tau=0$ and
\bez
\frac{d}{d\tau}|_{\tau = 0} {\cal  U}^{[\overline{X}]}_{h(\tau)}  \Psi
\in D([\overline{X}]).
\eez
}

{\bf Proof.}   It   is   sufficient  to  consider  the  case  $\Psi  =
v_{[\overline{X}]} \Phi$,          $\alpha          \in          {\cal
C}_0^{\infty}([\overline{X}])$, $\Phi \in D_0([\overline{X}])$. One has
\beb
\frac{{\cal U}^{[\overline{X}]}  - 1}{\tau} v_{[\overline{X}]}[\alpha]
\Phi = \tau^{-1} \{
v_{[\overline{X}]} [w_{h(\tau)}\alpha]                           {\cal
U}^{[\overline{X}]}_{h(\tau)} - v_{[\overline{X}]}[\alpha]
\} \Phi = \\
v_{[\overline{X}]} [\frac{w_{h(\tau)}\alpha - \alpha}{\tau}]
{\cal U}^{[\overline{X}]}_{h(\tau)} \Phi +
v_{[\overline{X}]} [\alpha] \frac{{\cal  U}^{[\overline{X}]}_{h(\tau)}
- 1}{\tau} \Phi.
\l{16}
\eeb
Since
\bez
\sup_{X \in [\overline{X}]}|
\frac{w_{h(\tau)}\alpha - \alpha}{\tau} -
\frac{d}{d\tau}|_{\tau = 0} w_{h(\tau)} \alpha)_X|
= \int_0^1 d\alpha \sup_{\overline{g} \in G/G_0(\overline{X})}
[\frac{d}{dt}|_{t=\alpha\tau} -        \frac{d}{d\tau}|_{t=0}        ]
\alpha_{u_{h^{-1}(\tau) \overline{g}}[\overline{X}]} \to_{\tau\to 0} 0,
\eez
expression \r{16} is strongly convergent to
\bez
v_{[\overline{X}]}[\frac{d}{d\tau}|_{\tau=0} w_{h(\tau)}\alpha] \Phi +
v_{[\overline{X}]}[\alpha] \frac{d}{d\tau}|_{\tau    =    0}     {\cal
U}^{[\overline{X}]}_{h(\tau)} \Phi \in D([\overline{X}]).
\eez
Lemma 3.4 is proved.

Let $h_A(\tau) = \exp A\tau$ be a one-parametric subgroup of the group
$G$, $A$ be its generator being an element of  the  corresponding  Lie
algebra. By  $\check{H}(A):  D([\overline{X}])  \to D([\overline{X}])$
denote the operator of the form
\beq
\check{H}(A) =        i        \frac{d}{d\tau}|_{\tau=0}         {\cal
U}^{[\overline{X}]}_{\exp A\tau}
\l{16a}
\eeq
Note that the operator $\check{H}(A)$ may be defined on the set of all
sections $\Psi$ such that ${\cal U}^{[\overline{X}]}_{\exp A\tau}\Psi$
is strongly differentiable at $\tau = 0$.

{\bf Lemma 3.5.} {\it  For  any  $\Psi  \in  D_0([\overline{X}])$  the
following properties are satisfied:
\beb
\check{H}(A_1+A_2) \Psi  =  \check{H}(A_1) \Psi + \check{H}(A_2) \Psi;
\qquad
\check{H}(\lambda A) \Psi = \lambda \check{H}(A) \Psi;\\
{\cal U}^{[\overline{X}]}_h             \check{H}(A)             {\cal
U}^{[\overline{X}]}_{h^{-1}} \Psi = \check{H}(hAh^{-1}) \Psi;
\qquad
[\check{H}(A); \check{H}(B)] \Psi = i \check{H}([A;B]) \Psi.
\l{18}
\eeb

{\bf Proof.}  By  $\nabla(A)$  we denote the following operator in the
space of smooth  functions  $\gamma:  G  \to  {\bf  C}$  with  compact
supports:
\bez
(\nabla(A) \gamma)(g)  =  \frac{d}{d\tau}|_{\tau=0}  \gamma(\exp A\tau
\cdot g).
\eez
}
It follows from eq.\r{15} that
\beq
\check{H}(A) \int     d_Lg     \gamma(g)    U_g(X\gets    u_{g^{-1}}X)
\Phi_{u_{g^{-1}}X} =  \int  d_L  g  [-i\nabla(A)\gamma](g)  U_g(X\gets
u_{g^{-1}}X) \Phi_{u_{g^{-1}}X}.
\l{17}
\eeq
Let su check now properties \r{18}.  The first relation is a corollary
of linearity of $\nabla(A)$ with respect to $A$:
\bez
\nabla(A_1+A_2) = \nabla(A_1) + \nabla(A_2); \qquad
\nabla(\lambda A) = \lambda \nabla(A).
\eez
To check the second property, introduce the operator
\bez
(W_h \gamma)(g) = \gamma(h^{-1}g).
\eez
Notice that
\beq
W_h \nabla(A) W_{h^{-1}} = \nabla(hAh^{-1}).
\l{19}
\eeq
For section \r{13}, one has
\bey
({\cal U}_h^{[\overline{X}]}             \check{H}(A)            {\cal
U}_{h^{-1}}^{[\overline{X}]} \Psi)_X = \int d_Lg
[-iW_h \nabla(A) W_{h^{-1}} \gamma](g) U_g(X \gets u_{g^{-1}}X)
\Phi_{u_{g^{-1}}X};\\
(\check{H}(hAh^{-1}) \Psi)_X = \int d_L g
[-i\nabla(hAh^{-1})\gamma)(g)
U_g(X \gets u_{g^{-1}}X)
\Phi_{u_{g^{-1}}X}.
\eey
The second property is proved.

Let $h(t) = \exp Bt$.  Differentiate eq.\r{19} with respect to $t$  at
$t=0$. We obtain:
\beq
[\nabla(A);\nabla(B)] + \nabla([A;B]) = 0.
\l{20}
\eeq
This implies the last relation. Lemma 3.5 is proved.

Consider the    following    operator    in    ${\cal    C}^{\infty}_0
([\overline{X}])$:
\bez
(\delta[A]\alpha)_X = \frac{d}{d\tau}|_{\tau=0} \alpha_{u_{\exp A\tau}
X}.
\eez

{\bf Lemma 3.6.} {\it  For $\Psi \in D(([\overline{X}]))$,
\beq
i[\check{H}(A); v_{[\overline{X}]} [\alpha]] \Psi =
v_{[\overline{X}]} [\delta[A]\alpha] \Psi.
\l{20a}
\eeq
}

The proof is straightforward.

{\bf Remark.}  Eq.\r{20a}  is  a  mathematical  formulation   of   the
heuristic fact that
\bez
\check{H}(A) = H(A:X) - i\delta[A]
\eez
for some operator $H(A:X)$ acting in ${\cal F}_X$.

{\bf Lemma  3.7.}  {\it  Properties \r{18} are satisfied for $\Psi \in
D([\overline{X}])$.}

{\bf Proof.} It is sufficient to check that relations \r{18} are valid
for sections   $v_{[\overline{X}]}   [\alpha]\Psi$   for   $\Psi   \in
D_0([\overline{X}])$, $\alpha \in {\cal C}_0^{\infty}([\overline{X}])$.
Combining Lemmas 2.5 and 3.6, we obtain
\bey
i (\check{H}(A_1+    \lambda A_2)   -    \check{H}(A_1)    -   \lambda
\check{H}(A_2)) v_{[\overline{X}]}[\alpha] \Psi =
v_{[\overline{X}]} [(\delta[A_1+\lambda  A_2]  -\delta[A_1]  - \lambda
\delta[A_2]) \alpha) \Psi = 0;\\
{\cal U}^{[\overline{X}]}_h             \check{H}(A)             {\cal
U}^{[\overline{X}]}_{h^{-1}} v_{[\overline{X}]} [\alpha] \Psi =
{\cal U}_h^{[\overline{X}]}                               \check{H}(A)
v_{[\overline{X}]}[w_{h^{-1}}\alpha]                             {\cal
U}^{[\overline{X}]}_{h^{-1}} \Psi  \\
=   {\cal   U}_h^{[\overline{X}]}
v_{[\overline{X}]}[w_{h^{-1}}\alpha] \check{H}(A)                {\cal
U}_{h^{-1}}^{[\overline{X}]} \Psi - i {\cal U}_h^{[\overline{X}]}
v_{[\overline{X}]}[\delta[A]w_{h^{-1}}\alpha]                    {\cal
U}^{[\overline{X}]}_{h^{-1}} \Psi = \\
v_{[\overline{X}]}[\alpha] \check{H}(hAh^{-1}) \Psi -
i v_{[\overline{X}]}[w_h \delta[A] w_{h^{-1}} \alpha] \Psi;\\
\check{H}(hAh^{-1}) v_{[\overline{X}]}[\alpha] \Psi =
v_{[\overline{X}]} [\alpha] \check{H}(hAh^{-1}) \Psi -
iv_{[\overline{X}]} [\delta[hAh^{-1}] \alpha] \Psi.
\eey
The property $w_h\delta[A]w_{h^{-1}} =  \delta[hAh^{-1}]$  is  checked
analogously to eq.\r{19}.

To check  the third property of set \r{18},  it is sufficient to check
that
\bez
[[\check{H}(A); \check{H}(B)]; v_{[\overline{X}]}[\alpha]] =
i[\check{H}([A;B]); v_{[\overline{X}]}[\alpha]].
\eez
This is equivalent to
\bez
[\delta[A];\delta[B]] + \delta[A;B] = 0.
\l{20b}
\eez
Relation \r{20b} is justified analogously to eq.\r{20}.  Lemma 3.7  is
proved.

{\bf Lemma      3.8.}      {\it     Let     the     section     ${\cal
U}^{[\overline{X}]}_{h(t)}\Psi$ be strongly differentiable  at  $t=0$.
Then the  differential operator $\delta[A]$ is defined on the function
$<\Psi,\Psi>_X$ and
\beq
-i\delta[A] <\Psi,\Psi>    =    <\Psi,    \check{H}(A)     \Psi>     -
<\check{H}(A)\Psi, \Psi>.
\l{21}
\eeq
}

{\bf Proof.} One has
\bey
\frac{1}{\tau} (<\Psi,\Psi>_{u_{\exp(-A\tau)}X} - <\Psi,\Psi>_X) =
\frac{1}{\tau} (<{\cal  U}^{[\overline{X}]}_{\exp  A\tau}\Psi,   {\cal
U}^{[\overline{X}]}_{\exp A\tau}\Psi>_X - <\Psi,\Psi>_X) =\\
(( \frac{{\cal U}^{[\overline{X}]}_{\exp A\tau} - 1}{\tau} \Psi)_X,
({\cal U}^{[\overline{X}]}_{\exp A\tau} \Psi)_X) +
(\Psi_X, (\frac{{\cal U}^{[\overline{X}]}_{\exp A\tau} - 1}{\tau}
\Psi)_X).
\eey
As $\tau\to 0$, this expression tends to
\bez
-i (\Psi_X, (\check{H}(A) \Psi)_X) + i ((\check{H}(A) \Psi)_X, \Psi_X).
\eez
Lemma 3.8 is proved.

Results of this section can be formulated as follows.

{\bf Theorem 3.1.} {\it Let the group $G$
act on the semiclassical bundle $({\cal Z},{\cal  X},\pi)$.  Then  the
operators $\check{H}(A):  D(A)  \to  D(A)$  of  the  form  \r{16a} are
defined an   the    dense    subset    $D({[\overline{X}]})    \subset
C({[\overline{X}]})$. The  properties  \r{18},  \r{20a} and \r{21} are
satisfied.
}

\section{Sufficient conditions    of    integrability    of    algebra
representation}

Starting from known operators ${\cal U}^{[\overline{X}]}_g$,  we  have
constructed the  generators  $\check{H}(A)$ of the representation.  An
inverse problem is  considered  in  this  section:  we  are  going  to
reconstruct operators   ${\cal   U}^{[\overline{X}]}_g$   from   known
$\check{H}(A)$.

Let us impose the following conditions  ("axioms")  on  the  operators
$\check{H}(A)$.

{\bf A1.}   {\it   There   exists   such   set   $D$  being  dense  in
$C({[\overline{X}]})$ such  that  the  operators  $\check{H}(A)$   are
defined on $D$.}

{\bf A2.} {\it
For any $\Phi,\Psi \in D$ the function $<\Phi,\Psi>$  belongs  to  the
domain of $\delta[A]$ and
\bez
-i\delta[A] <\Phi,\Psi>  =  <\Phi,  \check{H}(A) \Psi> - <\check{H}(A)
\Phi,\Psi>.
\eez
}

{\bf A3.}   {\it   For   any   $\alpha   \in    {\cal    C}_0^{\infty}
({[\overline{X}]})$ and  $\Psi \in D$ $v_{[\overline{X}]}[\alpha] \Psi
\in D$ and
\bez
[\check{H}(A); v_{[\overline{X}]}[\alpha]]       \Psi       =        -
iv_{[\overline{X}]} [\delta[A] \alpha].
\eez
}

{\bf A4.}    {\it    For    $\Phi,\Psi    \in    D$    the    function
$<\Phi,\check{H}(A)\Psi>$ belongs to the domain of $\delta[B]$.
}

{\bf A5.} {\it For $\Phi,\Psi \in D$ the property
\bey
<\check{H}(A)\Psi; \check{H}(B)\Phi> - i \delta[A] <\Psi, \check{H}(B)
\Phi> - <\check{H}(B) \Psi,  \check{H}(A) \Phi>  +  i\delta[B]  <\Psi,
\check{H}(A)\Phi> \\
= i <\Psi, \check{H}([A;B]) \Phi>.
\eey
}

{\bf A6.}  {\it  Let  $B_1,...,B_n$  be  a  basis  in the Lie algebra.
Suppose that $\Psi^0\in D$.  Then there exists a solution $\Psi^t$  of
the Cauchy problem for the equation
\beq
i\frac{d}{dt} \Psi^t = \check{H}(B_k) \Psi^t
\l{22}
\eeq
such that  $\Psi^t$  is  strongly  continuously  differentiable   with
respect to $t$. Moreover,
\bez
\check{H}(B_l) \Psi^t \to \check{H}(B_l) \Psi^0
\eez
as $t\to 0$ in the strong sense for all $l$, while the function
\bez
\delta[A]<\Phi,\check{H}(B_k) \Psi_t>
\eez
is continuous.
}

Under conditions A1-A6,  it is possible to construct an action of  the
Lie group $G$ on the semiclassical bundle.

{\bf Lemma 4.1.} {\it The solution $\Psi^t$ of eq.\r{22} satisfies the
following property
\bez
<\Psi^t,\Psi^t>_X = <\Psi^t,\Psi^t>_{u_{\exp (-B_kt)}X}.
\eez
}

{\bf Proof.} Consider the function $f_X(t) =  <\Psi^t,\Psi^t>_X$.  One
has
\bey
\frac{1}{\tau} (f_X(t+\tau) - f_X(t)) =
<\frac{\Psi^{t+\tau} - \Psi^t}{\tau}; \Psi^{t+\tau}>_X
+ <\Psi^t, \frac{\Psi^{t+\tau} - \Psi^t}{\tau}>_X \to_{\tau\to 0}\\
<-i\check{H}(B_k) \Psi^t,  \Psi^t>_X  +  <\Psi^t,  -   i\check{H}(B_k)
\Psi^t>_X = - (\delta[B_k] <\Psi^t,\Psi^t>)_X = - \delta[B_k] f_X(t).
\eey
Hence, $\frac{d}{dt} f_{u_{\exp B_kt}X} = 0$. Lemma 4.1 is proved.

{\bf Corollary.}  There exists a unique solution of the Cauchy problem
for eq.\r{22}.

{\bf Proof.} Assume the converse, i.e. $\Psi^0=0$, $\Psi^t\ne 0$. This
contradicts Lemma 4.1. Corollary is proved.

By ${\cal  U}^t_{B_k}$  we denote the operator that takes each initial
condition $\Psi^0$ for the Cauchy problem to the solution $\Psi^t$  to
the Cauchy problem for eq.\r{22},
\bez
U^t_{B_k} \Psi^0 = \Psi^t.
\eez
Combining Lemma 4.1 and axiom A6, we obtain:

{\bf Lemma 4.2.} {\it The operator $U^t_{B_k}$ satisfies the following
properties:
\bez
U^t_{B_k}: D \to D; \qquad
U^{t+\tau}_{B_k} = U^t_{B_k} U^{\tau}_{B_k}
\eez
for any $\Psi \in D$;
\bez
i \frac{U^t_{B_k} -1}{t} \Psi \to_{t\to 0} \check{H}(B_k) \Psi;
\qquad
\check{H}(B_s) U^t_{B_k} \Psi \to_{t\to 0} \check{H}(B_s) \Psi
\eez
in the strong sense for any $\Psi \in D$;
\bez
<U^t_{B_k} \Psi, U^t_{B_k} \Psi>_X = <\Psi,\Psi>_{u_{\exp(-B_kt)}X}.
\eez
}

The proof is obvious.

{\bf Corollary.} {\it The operator $U^t_{B_k}$ is isometric. It can be
uniquely extended to $C({[\overline{X}]})$.
}

{\bf Lemma 4.3.} {\it The following property is satisfied:
\bez
U^t_{B_k} v_{[\overline{X}]}[\alpha]   =  v_{[\overline{X}]}  [w_{\exp
B_kt} \alpha] U^t_{B_k}.
\eez
}

{\bf Proof.} Consider the following initial condition for  the  Cauchy
problem for eq.\r{22}:
\bez
\Phi_X^0 = \alpha_X \Psi_X^0.
\eez
To prove Lemma 4.3, it is sufficient to justify that section
\bez
\Phi^t_X = \alpha_{u_{\exp (-B_kt)} X} \Psi_X^t
\eez
satisfies eq.\r{22}. One has:
\bey
\frac{i}{\tau} (\Phi_X^{t+\tau} - \Phi_X^t) = i
\frac{
\alpha_{u_{\exp(-B_k(t+\tau))}X}
- \alpha_{u_{\exp(-B_kt)}X}}{\tau} \Psi^{t+\tau}_X +
i \alpha_{u_{\exp(-B_kt)}X} [\frac{\Psi_X^{t+\tau} - \Psi_X^t}{\tau}].
\eey
Consider the strong limit $\tau\to 0$. Then we get
\bez
i \frac{d}{dt}   \Phi_X^t  =  i(\delta[B_k]\alpha)_{u_{\exp(-B_kt)}X}
\Psi^t_X + \alpha_{u_{\exp(-B_kt)}X} (\check{H}(B_k) \Psi^t)_X =
(\check{H}(B_k) \alpha_{u_{\exp(-B_kt)}X} \Psi^t)_X = (\check{H}(B_k)
\Phi^t)_X.
\eez
Lemma 4.3 is proved.

{\bf Lemma 4.4.} {\it For $\Psi \in D$,
\bez
U^{-t}_{B_k} \check{H}(A) U^t_{B_k} \Psi =
\check{H}(e^{-B_kt} A e^{B_kt}) \Psi.
\eez
}

{\bf Proof.} It is sufficient to check that the function
\bez
f_X^t =  <U^t_{B_k}  \Phi,  \check{H}(e^{B_kt}  A e^{-B_kt}) U^t_{B_k}
\Psi>_X, \qquad \Phi,\Psi \in D.
\eez
satisfies the property
\beq
f^t_X = f^0_{u_{e^{-B_kt}X}}.
\l{23}
\eeq
Consider the derivative of $f$ with respect to $t$. One has
\bey
\frac{1}{\tau} (f_X^{t+\tau} - f_X^t) = <U^{t+\tau}_{B_k} \Phi,
\check{H}(\frac{e^{B_k(t+\tau)} A  e^{-B_k(t+\tau)}   -   e^{B_kt}   A
e^{-B_kt} }{\tau}) U^{t+\tau}_{B_k} \Psi>_X
+ <\frac{U^{\tau}_{B_k}  -  1}{\tau}  U^t_{B_k}   \Phi,   \check{H}(A)
U^{t+\tau}_{B_k} \Psi>_X \\
+ <U^t_{B_k} \Phi, \check{H}(A)
\frac{U^{\tau}_{B_k} - 1}{\tau} U^t_{B_k} \Psi>_X.
\eey
The first term contains the operator of the type
\bez
\check{H}(\sum_s c_{k,s}(\tau) B_s)
\eez
with
\bez
\sum_s c_{k,s} (\tau) B_s \to_{\tau\to 0}
[B_k; e^{B_kt} A e^{-B_kt}].
\eez
Therefore, it tends to
\bez
<U^t_{B_k} \Phi, \check{H}([B_k;e^{B_kt}Ae^{-B_kt}]) U^t_{B_k} \Psi>_X
\eez
The second term tends to
\bez
<-i\check{H}(B_k) U^t_{B_k} \Phi, \check{H}(A) U^t_{B_k} \Psi>_X.
\eez
The third term can be rewritten as a sum
\bez
<\check{H}(A) U^t_{B_k} \Phi, \frac{U^{\tau}_{B_k} -1}{\tau} U^t_{B_k}
\Psi>_X - i (\delta[A] <U^t_{B_k} \Phi, \frac{U^{tau}_{B_k} - 1}{\tau}
U^t_{B_k} \Psi>)_X.
\eez
The first term tends to
\bez
<\check{H}(A) U^t_{B_k} \Phi, -i\check{H}(B_k) U^t_{B_k} \Psi>_X,
\eez
the second term is taken to the form
\bez
-i\delta[A] <U^t_{B_k}   \Phi,   \int_0^1   d\alpha  (-i\check{H}(B_k)
\Psi^{t+\tau\alpha})>_X
\eez
and tends to
\bez
-\delta[A] <U^t_{B_k}\Phi, \check{H}(B_k) U^t_{B_k} \Psi>_X.
\eez
Combining all the terms and taking into account axiom A5, we find
\bez
\frac{d}{dt} f^t_X = - \delta[B_k] f^t_X.
\eez
This equation can be rewritten as
\bez
\frac{d}{dt} f^t_{u_{\exp B_kt}X} = 0.
\eez
Property \r{23} is checked. Lemma 4.4 is proved.

{\bf Lemma 4.5.} {\it Let $t_k(\alpha)$, $k=1,...,m$,
be smooth  functions  such that $t_k(0)=0$.  Suppose also that for any
$\alpha \in [0,1]$
\bez
e = \exp B_{i_1}t_1(\alpha) ... \exp B_{i_m}t_m(\alpha)
\eez
Then
\bez
U_{B_{i_1}}^{t_1(\alpha)} ... U_{B_{i_m}}^{t_m(\alpha)} = 1.
\eez
}

{\bf Proof.}  Consider the expression
\bey
i \frac{d}{d\alpha} (U_{B_{i_1}}^{t_1(\alpha)} ... U_{B_{i_m}}^{t_m(\alpha)})
(U_{B_{i_1}}^{t_1(\alpha)} ... U_{B_{i_m}}^{t_m(\alpha)})^{-1} = \\
\sum_{j=1}^m U_{B_{i_1}}^{t_1(\alpha)} ... U_{B_{i_{j-1}}}^{t_{j-1}(\alpha)}
i \frac{d}{d\alpha} U_{B_{i_j}}^{t_j(\alpha)} \cdot
U_{B_{i_j}}^{-t_j(\alpha)} ... U_{B_{i_1}}^{-t_1(\alpha)}=\\
\sum_{j=1}^m t_j'(\alpha)
U_{B_{i_1}}^{t_1(\alpha)} ... U_{B_{i_{j-1}}}^{t_{j-1}(\alpha)}
\check{H}(B_{i_j})
U_{B_{i_{j-1}}}^{-t_{j-1}(\alpha)} ... U_{B_{i_1}}^{-t_1(\alpha)}.
\eey
By Lemma 4.4, we take this expression to the form
\beb
\sum_{j=1}^m \check{H} (t_j'(\alpha) e^{B_{i_1} t_1(\alpha)}
... e^{B_{i_{j-1}} t_{j-1}(\alpha)} B_{i_j}
e^{- B_{i_{j-1}} t_{j-1}(\alpha)} ... e^{- B_{i_1} t_1(\alpha)})
\l{24}
\eeb
The property
\bez
\frac{d}{d\alpha} (e^{B_{i_1} t_1(\alpha)} ... e^{B_{i_m} t_m(\alpha)})
(e^{B_{i_1} t_1(\alpha)} ... e^{B_{i_m} t_m(\alpha)})^{-1} = 0
\eez
implies that expression \r{24} vanishes. Therefore,
\bez
U_{B_{i_1}}^{t_1(\alpha)} ... U_{B_{i_m}}^{t_m(\alpha)} = const = 1.
\eez
Lemma 4.5 is proved.

Let us construct now the operators ${\cal  U}_g^{[\overline{X}]}$.  It
is well known that one can locally introduce the second-kind canonical
coordinates $t_1,..,t_n$ on the Lie group $G$ as follows \c{P};
\beq
g = \exp B_1t_1 ... \exp B_nt_n.
\l{25}
\eeq
To each  $g$  of  the form \r{25} assign the following operator ${\cal
U}_g^{[\overline{X}]}$:
\beq
{\cal U}_g^{[\overline{X}]} = U^{t_1}_{B_1} ... U^{t_n}_{B_n}.
\l{26}
\eeq

{\bf Lemma 4.6.}  {\it  The  operators  ${\cal  U}_g^{[\overline{X}]}$
satisfy the following properties:\\
(i) the  operator  ${\cal  U}_g^{[\overline{X}]}$  is  isometric  with
respect to the $C({[\overline{X}]})$-norm;\\
(ii) the group relation \r{7a} is satisfied;\\
(iii) formula \r{10} is obeyed;\\
(iv) the operator ${\cal U}_g^{[\overline{X}]}$ is strongly continuous
with respect to $g$ at $g=e$.
}

{\bf Proof.}
Corollary of  Lemma  4.2  implies  the  isometric property.  Lemma 4.3
implies relation \r{10}.

Let us  check  that  the  operator  ${\cal  U}_g^{[\overline{X}]}$  is
strongly continuous with respect to $g$ at $g=e$.  Let $\{g_l \in G\}$
be any sequence that tends to $e$ as $l\to\infty$. Then
\bez
g_l = \exp B_1t_1^l ... \exp B_nt_n^l,
\eez
where $t_i^l \to 0$ as $l\to\infty$.  Since the operators  $U^t_{B_k}$
are strongly continuous, one has
\bez
{\cal U}_g^{[\overline{X}]} = U_{B_1}^{t_1^l} ...  U_{B_n}^{t_n^l} \to
1
\eez
in the strong sense.

Let us justify the group property \r{7a}.  For $g=e$,  one has $t_1  =
... = t_n = 0$ and ${\cal U}_g^{[\overline{X}]} = 1$. Let $g^1,g^2 \in
G$ and
\bez
g^{1,2} = \exp B_1t_1^{1,2} ... \exp B_n t_n^{1,2}.
\eez
Consider the functions
\bez
g^{1,2}(\alpha) = \exp (\alpha  B_1t_1^{1,2})  ...  \exp  (\alpha  B_n
t_n^{1,2})
\eez
and
\bez
g^1(\alpha)  g^2(\alpha) = \exp (B_1t_1^{3}(\alpha))
...  \exp  (B_n t_n^{3}(\alpha))
\eez
Then
\bez
e^{B_1t_1^1\alpha} ... e^{B_nt_n^1\alpha}
e^{B_1t_1^2\alpha} ... e^{B_nt_n^2\alpha}
e^{- B_nt_n^3(\alpha)} ... e^{B_1t_1^3(\alpha)} = e.
\eez
Lemma 4.5 implies that
\bez
U_{B_1}^{t_1^1\alpha} ... U_{B_n}^{t_n^1\alpha}
U_{B_1}^{t_1^2\alpha} ... U_{B_n}^{t_n^2\alpha}
U_{B_n}^{- t_n^3(\alpha)} ... U_{B_1}^{t_n^1(\alpha)} = 1.
\eez
This proves the group relation and Lemma 4.6.

The obtained results can be formulated as follows.

{\bf Theorem 4.1.} {\it Suppose that for any
$g\in G$ the transformation $u_g: {\cal X} \to {\cal X}$ such that\\
(i) the mapping $(g,X) \mapsto u_gX$ is smooth;\\
(ii) $u_{g_1g_2} = u_{g_1} u_{g_2}$\\
is specified. Suppose also that for any $[\overline{X}]$ the operators
$\check{H}(A)$ satisfying axioms A1-A6 are given.  Then there exists a
unique action of the local group $G$ on the semiclassical bundle  such
that
\beq
i \frac{d}{d\tau}|_{\tau=0} {U}_{g(\tau)} (X \gets u_{g^{-1}(\tau)} X)
\Psi_{u_{g^{-1}(\tau)} X} = (\check{H}(A) \Psi)_X
\l{27}
\eeq
for $g(\tau) = \exp A\tau$, $\Psi \in D$.
}

{\bf Proof.} Since the second-kind canonical coordinates  are  unique,
the operators  ${\cal  U}_g^{[\overline{X}]}$ \r{26} are well defined.
Differentiating formula \r{26} with respect to $\tau$ for  any  smooth
curve $g(\tau)$ on $G$, we obtain relation \r{26}.

\section{Gauge theories}

Let us generalize the obtained results to the gauge theories.

Let $\cal  L$  be  a  gauge  group  acting on the semiclassical bundle
$({\cal Z},{\cal X},\pi)$ (Definition 1.2).  In particular,  the group
$\cal L$ acts on the manifold $\cal X$.

{\bf Definition  5.1.}  {\it We say that a Lie group $G$ {\sf acts} on
the smooth manifold $\cal X$ with the gauge group $\cal L$ if for  any
$g\in G$ the transformation $u_g: {\cal X} \to {\cal X}$ such that \\
(i) the mapping $(g,X) \mapsto u_gX$ is smooth; \\
(ii) for  any $\alpha \in {\cal L}$,  $g\in G$ there exists $\beta \in
{\cal L}$  such  that  the  mapping  $(g,\alpha)  \mapsto  \beta$   is
continuous and
\beq
u_g \lambda_{\alpha} u_{g^{-1}} = \lambda_{\beta};
\l{28}
\eeq
(iii) for any $g_1,g_2 \in G$ there exists $\gamma \in {\cal L}$  such
that the mapping $(g_1,g_2) \mapsto \gamma$ is continuous and
\beq
u_{g_1} u_{g_2} = \lambda_{\gamma} u_{g_1g_2}
\l{29}
\eeq
is given.
}

Two points  $X_1,X_2 \in {\cal X}$ are called gauge-equivalent if $X_2
= \lambda_{\alpha} X_1$ for  some  $\alpha  \in  {\cal  L}$.  Property
\r{28} means  that  gauge-equivalents  points of $\cal X$ are taken to
gauge-equivalent. Property    \r{29}    means    that    the    points
$u_{g_1}u_{g_2} X$ and $u_{g_1g_2}X$ are gauge-equivalent.  This is an
analog of the group property.

{\bf Definition 5.2.} {\it We say that a Lie group $G$ {\sf acts}
on the semiclassical bundle $({\cal Z},{\cal X},\pi)$ if:
\\
(i) $G$ acts on the manifold $\cal X$;\\
(ii) for any $g\in G$ and $X \in {\cal X}$ the unitary operators
$U_g(u_gX \gets X): {\cal F}_X \to {\cal F}_{u_gX}$ such that
\\
(1) the mapping $(g,h,\alpha) \mapsto U_g(u_h\lambda_{\alpha} X \gets
u_{g^{-1}}u_h\lambda_{\alpha} X)$ is strongly continuous with  respect
to $(g,h,\alpha)$;\\
(2) the following relations are satisfied:
\beq
U_g(u_g\lambda_{\alpha}X \gets      \lambda_{\alpha}X)      V_{\alpha}
(\lambda_{\alpha} X\gets  X)  U_{g^{-1}}  (X\gets  u_gX)  =  V_{\beta}
(u_g\lambda_{\alpha}X \gets u_gX);
\l{30}
\eeq
\beq
U_g(u_{g_1}u_{g_2}X \gets  u_{g_2}X)  U_{g_2}  (u_{g_2}X  \gets  X)  =
V_{\beta} (\lambda_{\beta}  u_{g_1g_2} X \gets u_{g_1g_2}X) U_{g_1g_2}
(u_{g_1g_2}X \gets X)
\l{31}
\eeq
are given.
}

Let us construct an analog of the space $C([\overline{X}])$. By
$(Z_{[\overline{X}]},{[\overline{X}]},\pi)$ we denote the subbundle of
the form ${[\overline{X}]} = \{ \lambda_{\alpha}u_g {[\overline{X}]} |
g\in G,    \alpha    \in    {\cal    L}\}$,    $Z_{[\overline{X}]}   =
\pi^{-1}{[\overline{X}]}$. By $G_0({\overline{X}})$ denote the set  of
all $h\in       G$       such      that      $u_h{\overline{X}}      =
\lambda_{\alpha}{\overline{X}}$ for some $\alpha \in {\cal L}$. We say
that a strongly continuous section $\Psi_Y$ of the subbundle
$Z_{[\overline{X}]},{[\overline{X}]},\pi)$ is     of     the     class
${C}_0({[\overline{X}]})$ if    it    is    invariant    under   gauge
transformations,
\beq
\Psi_{\lambda_{\alpha}Y} =  V_{\alpha}  (\lambda_{\alpha}Y  \gets   Y)
\Psi_Y,
\l{32}
\eeq
while the     set     of     all     classes     $\overline{g}     \in
G/G_0({[\overline{X}]})$ such                                     that
$\Psi_{\lambda_{\alpha}u_{\overline{g}} {[\overline{X}]}}  \ne  0$  is
pre-compact in $G/G_0({[\overline{X}]})$. Introduce the following norm
in $C_0({[\overline{X}]})$:
\beq
||\Psi||_{[\overline{X}]} = \sup_{X \in {[\overline{X}]}} ||\Psi_X||.
\l{33}
\eeq
By $C({[\overline{X}]})$      we     denote     the     closure     of
$C_0({[\overline{X}]})$ with respect to the norm \r{33}.  Consider the
operator
\beq
({\cal U}^{[\overline{X}]}_g   \Psi)_Y   =  U_g(Y  \gets  u_{g^{-1}}Y)
\Psi_{u_{g^{-1}}Y}.
\l{34}
\eeq

{\bf Theorem 5.1.} {\it  Let the Lie group $G$ act on the semiclassical
bundle $({\cal  Z},{\cal X},\pi)$ with the gauge group $\cal L$.  Then
the operator \r{34}:\\
(i) maps $C_0({[\overline{X}]})$ to $C_0({[\overline{X}]})$;\\
(ii) conserves the norm \r{33};\\
(iii) can be uniquely extended to $C({[\overline{X}]})$;\\
(iv) satisfies the group property \r{7a};\\
(v) is strongly continuous with respect to $g$ at $g=e$.\\
Property \r{10} is satisfied for gauge-invariant  functions  $\alpha$,
which satisfy the property
\beq
\alpha_{\lambda_{\gamma}Y} =  \alpha_Y;  \qquad  {\rm  for \qquad all}
\qquad \gamma \in {\cal L}.
\l{34a}
\eeq
}

The proof is analogous to Theorem 2.1.

{\bf Theorem 5.2.} {\it
Let the Lie group $G$ act on the manifold  $\cal  X$  with  the  gauge
group $\cal   L$.  Let  the  operators  ${\cal  U}^{[\overline{X}]}_g:
C({[\overline{X}]}) \to C({[\overline{X}]})$ such that:\\
(i) ${\cal U}^{[\overline{X}]}_g$ conserves the norm;\\
(ii) ${\cal U}^{[\overline{X}]}_g$ is strongly continuous with respect
to $g$ at $g=e$;\\
(iii) the group property \r{7a} is obeyed;\\
(iv) relation  \r{10}  is satisfied for all functions $\alpha$ obeying
eq.\r{34a};\\
be given.  Then there exists a unique set of operators $U_g(u_gX \gets
X): {\cal F}_X \to {\cal F}_{u_gX}$,  $X\in {\cal X}$,  $g\in G$  such
that\\
(i) the mapping $(g,h,\alpha) \mapsto  U_g(u_h\lambda_{\alpha}X  \gets
u_{g^{-1}}u_h\lambda_{\alpha}X)$ is  strongly continuous with respect
to $(g,h,\alpha)$;\\
(ii) properties \r{30}, \r{31} and \r{34} are satisfied.
}

The proof is analogous to Theorem 2.2.

Define the   garding   domain   analogously   to   section    3.    By
$D_0({[\overline{X}]})$ we  denote the set of all sections of the type
\r{13} with $\Phi \in C_0({[\overline{X}]})$.  By ${\cal C}_0^{\infty}
({[\overline{X}]})$ we   denote   the  set  of  all  smooth  functions
$\alpha_X: {[\overline{X}]} \mapsto {\bf C}$ such that  eq.\r{34a}  is
satisfied and    the    set   of   all   classes   $\overline{h}   \in
G/G_0({\overline{X}})$ such       that       $\alpha_{u_{\overline{h}}
{\overline{X}}} \ne  0$  is  pre-compact.  By $D({[\overline{X}]})$ we
denote the set of all linear combinations
\bez
v_{[\overline{X}]}[\alpha_1] \Psi_1 + ... +
v_{[\overline{X}]}[\alpha_s] \Psi_s,
\qquad
\alpha_i \in {\cal C}_0^{\infty}({[\overline{X}]}), \qquad
\Psi_i \in D_0({[\overline{X}]}).
\eez

{\bf Theorem 5.3.} {\it Let the Lie group $G$ act on the semiclassical
bundle $({\cal Z},{\cal X},\pi)$ with the gauge group $\cal L$. Then:\\
(i) $D({[\overline{X}]})$ is a dense subset of $C({[\overline{X}]})$;\\
(ii) the operators $\check{H}(A)$ of the form \r{16a} are  defined  on
$D({[\overline{X}]})$;\\
(iii) $\check{H}(A): D({[\overline{X}]}) \to D({[\overline{X}]})$;\\
(iv) properties \r{18}, \r{20a}, \r{21} are satisfied.
}

The proof is analogous to Theorem 3.1.

{\bf Theorem 5.4.} {\it Let the Lie group  $G$  act  on  the  manifold
$\cal X$ with the gauge group $\cal L$. Let for any ${[\overline{X}]}$
the operator $\check{H}(A)$ in $C({[\overline{X}]})$ be given. Let the
axioms A1-A6  be  satisfied.  Then there exists a unique action of the
local Lie group on he bundle $({\cal Z}, {\cal X}, \pi)$ such that the
property \r{27}  for  $\Psi \in D({[\overline{X}]})$,  $g(\tau) = \exp
A\tau$ is satisfied.
}

The proof is analogous to Theorem 4.1.

\section*{Acknowledgments}

This work  was supported by the Russian Foundation for Basic Research,
projects 99-01-01198 and 01-01-06251.


\end{document}